\documentclass[aps,prb,showpacs,amsfonts,amssymb,amsmath,floatfix,twocolumn,
floats,nobalancelastpage]{revtex4}

\usepackage{graphicx}

\begin{document}          

\title{A Discrete-Event Analytic Technique for Surface Growth Problems}

\author{A. Kolakowska}
\email{alicjak@bellsouth.net}
\author{M. A. Novotny}
\email{man40@ra.msstate.edu}
\affiliation{Department of Physics and Astronomy, and the ERC Center for 
Computational Sciences,
P.O. Box 5167, Mississippi State, MS 39762-5167}

\date{\today}

\begin{abstract}
We introduce an approach for calculating non-universal properties of rough surfaces. 
The technique uses concepts of distinct surface-configuration 
classes, defined by the surface growth rule. The key idea is a mapping between 
discrete events that take place on the interface and its elementary local-site 
configurations. We construct theoretical probability distributions of deposition 
events at saturation for surfaces generated by selected growth rules. These 
distributions are then used to compute measurable physical quantities. Despite 
the neglect of temporal correlations, our approximate analytical results are in 
very good agreement with numerical simulations. This discrete-event analytic 
technique can be particularly useful when applied to quantification problems, 
which are known to not be suited to continuum methods.
\end{abstract}


\maketitle

\section{INTRODUCTION \label{intro}}


Broad interest in surface growth problems and the interface motion stems from 
many applications that these problems have in practically all sciences. 
Some well-known examples are crystallization problems, epitaxial growth, 
wetting phenomena, electrophoretic deposition of polymer chains, 
and growth of bacterial colonies. The mainstream studies of physical mechanisms 
that lead to growth and roughening, and interface properties, focus on the 
so-called universal properties, i.e., large-scale properties of growing 
surfaces as determined by universal scaling exponents. A central point here 
is a coarse-grained description of the interface, which leads to an effective 
growth equation, typical for a universality class. It is safe to say that 
universal properties of time-evolving surfaces are well understood \cite{BS95}. 
There are many situations, however, where {\it non-universal} properties, i.e., 
those pertaining to the microscopic structure of the interface, are of 
importance in understanding the physical system. One case is the density of 
local minima of a virtual-time interface, evolving in parallel discrete-event 
simulations \cite{KTNR00,KNGTR03,KNR03}. Another case is the movement of 
random walkers on a growing surface \cite{Ch02}. Another case is the density of 
maxima after solidification and processing of a surface, where this density 
is related to the frictional force \cite{SDG01}. Even for a one-dimensional 
interface, as in the growth of a two-dimensional crystal \cite{SJB00}, 
a general description of the microscopic structure is challenging, both 
for analytics and for simulations, due to a variety of length and time scales. 
However, for a strongly non-equilibrium interface (e.g., in a cold system) 
a net picture of the interface gains and losses can be simplified. 
This leads to a variety of  simulation models such as, e.g., ballistic 
deposition, Eden or solid-on-solid models.

Here we introduce a discrete-event analytic technique that provides a means 
for calculating some non-universal properties of interfaces. The key idea 
is the construction of a probability distribution for events that take place 
on the surface. This new approach gives a mean-field like approximation of  
averages that are determined in simulations. In Sec.~\ref{method}, we explain the main 
concept by examples of applications to selected non-equilibrium 
one-dimensional model interfaces. The method is summarized in Sec.~\ref{summary}.

\section{THE METHOD \label{method}}

\begin{figure}[b]
\includegraphics[width=7.0cm]{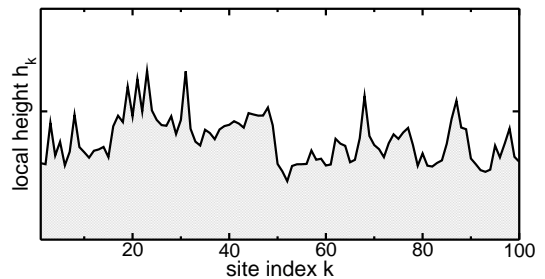}
\caption{\label{figure1} A typical model surface in {\it 1D}, $L=100$.}
\end{figure}

We focus on a model in which a one dimensional ({\it 1D}) surface grows by the 
deposition of local height increments $\eta_k$ at $L$ lattice sites, with periodic 
boundary condition $h_{L+1}=h_1$ ($k$ enumerates the lattice sites, $h_k$ is the local height). 
The deposited $\eta_k$ is a real positive number that can take on 
continuous values in the interval $[0;\infty)$. Our discussion is not specific to the 
probability distribution from which $\eta_k$ is sampled. We consider a rough surface at 
saturation (Fig.~\ref{figure1}). Such a surface can be generated, e.g., in the single-step 
solid-on-solid models \cite{xxx} or in random deposition with surface relaxation 
\cite{Fam86} or by depositing random $\eta_k$ at local surface 
minimum \cite{KTNR00,KNR03}.

In the latter case, the growth is simulated by the nearest-neighbor rule: 
$h_k(t+1)=h_k(t)+ \eta_k(t)$, if $h_k(t) \le \min \{ h_{k-1}(t), h_{k+1}(t)\}$; 
otherwise, $h_k(t+1)=h_k(t)$ ($t$ is the time index). The above 
deposition rule produces surfaces characterized by the universal roughness exponent 
\cite{KTNR00} $\alpha =1/2$. Our objective is to show that 
it is possible to construct a theoretical 
probability distribution of deposition events on the surface generated by the above 
rule. This distribution can then be used to compute measurable  
quantities at saturation, e.g., the mean density of local minima or maxima. Also, it can 
serve as the departure point for the construction of other distributions that describe 
more complex growth processes. Our derivation makes the following two simplifying 
assumptions. First, we neglect correlations between nearest-neighbor local slopes, 
which depend on the type of deposition, i.e., the distribution from which $\eta_k$ 
is sampled. Second, we neglect temporal correlations among the groups of the surface 
configuration classes, as explained later. Because of the above simplifications, 
our theoretical results for the averages are a mean-field like approximation 
to the averages measured in simulations.

\begin{figure}[!]
\includegraphics[width=7.5cm]{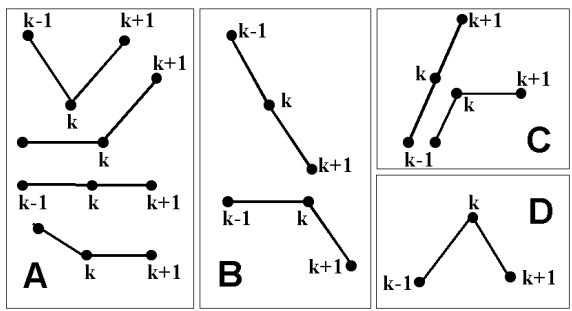}
\caption{\label{figure2} The four groups of elementary local surface
configurations at the $k$-th site. In accordance  with the adopted
growth rule, local A-configuration represents four types of discrete deposition
events, and local configuration B (or C) represents two types of events.}
\end{figure}

The adopted growth rule provides the principle 
for the classification of local lattice-site configurations of the interface. There are 
only four groups of these configurations (Fig.~\ref{figure2}). Each group corresponds 
to one of the four mutually exclusive discrete events A, B, C and D that take place at 
site $k$ during the deposition attempt at $t$: ``A" denotes an event when the 
deposition rule is satisfied from both sides, i.e., $h_k(t)\le h_{k-1}(t)$ and 
$h_k(t) \le h_{k+1}(t)$; ``B" denotes an event when the rule is not satisfied 
from the right; ``C" denotes 
an event when the rule is not satisfied from the left; ``D" denotes an event when 
the rule is not satisfied from either side. The above events can be mapped in 
unique way onto elementary local site configurations A, B, C and D, presented in 
Fig.~\ref{figure2}. For a closed chain of sites, the set of events when all 
sites are simultaneously in the A-configuration is of measure zero, and not all $L$ sites 
can have the same elementary configuration. Therefore, in the set of $L$ sites there must be at 
least one site with configuration A. We assign the index $k=1$ to one of the sites in 
the A-configuration and enumerate the other sites accordingly, progressing to the right. 
Its right neighbor (site $k=2$) can be only either in configuration C or D. Similarly, 
its left neighbor (site $k=L$) can be only either in B or D. If site $k=2$ is in C 
then site $k=3$ can be only either in C or D. If site $k=2$ is in D then site $k=3$ 
can be only either in B or A. If site $k=3$ is in B then its right neighbor can be 
either in B or A, etc. Starting from the site $k=1$ and progressing to the right 
towards $k=L$, applying these elementary neighbor rules, we can 
construct all possible configuration equivalency classes of the entire surface 
generated by the adopted growth rule. These can be categorized into groups 
(called $p$-groups), based on the number $p$ of the deposition events on the surface 
at $t$, i.e., here, the number of local minima in the surface configuration 
(coded by ``A") at $t$. The probability distribution $f(p; L)$ of the deposition 
events is obtained as the quotient of the multiplicity $M(p)$ of the $p$-group 
and the total number $M$ of configuration classes. 
(We omit $t$ in the notation since the analysis concerns  
steady-state evolution.) The mean density $\langle u(L) \rangle$ 
of local minima is given by the generally valid formula:
\begin{equation}
\label{average}
\langle u (L) \rangle = \sum_{p} u(p) f(p;L) \, ,
\end{equation}
where the summation is over all $p$-groups of the admissible surface-configuration 
classes, $u(p)$ is the density characteristic for each group, and $f(p;L)$ is the 
frequency of the occurrence of the $p$-group at saturation, i.e., the probability 
of generating a surface with $p$ local minima.

\begin{figure}[th!]
\includegraphics[width=7.0cm]{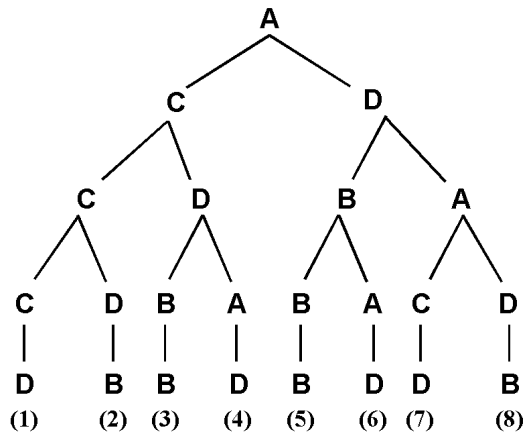}
\caption{\label{figure3} Binary tree for the construction of all possible configuration
classes for L=5.}
\end{figure}

\begin{figure}[th!]
\includegraphics[width=7.0cm]{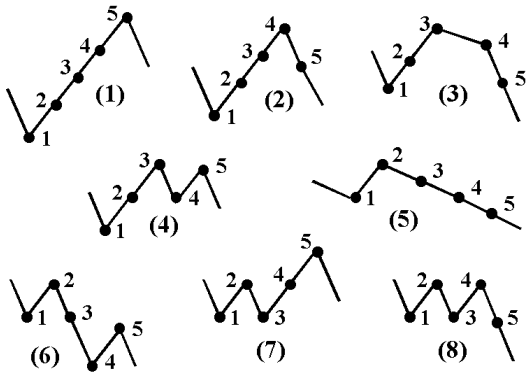}
\caption{\label{figure4} The graphs of possible surface-configuration classes, 
corresponding to the configurations read along the branches from Fig.~\ref{figure3} 
(see text). Each graph represents a class of infinitely many topologically 
equivalent deformations.}
\end{figure}

For example, the binary tree for the construction of possible surface-configuration 
classes for $L=5$ is shown in Fig.~\ref{figure3}. Looking along its branches, starting from 
the leading A at the fixed $k=1$ position, one identifies $M=8$ configuration classes of 
the entire surface: (1)~ACCCD; (2)~ACCDB; (3)~ACDBB; (4)~ACDAD; (5)~ADBBB; (6)~ADBAD; 
(7)~ADACD; and (8)~ADADB. Each surface configuration represents a class of infinitely many 
topologically equivalent deformations (Fig.~\ref{figure4}) since the deposited random 
height increment is a real positive number. There are only two $p$-groups. In the first, 
there are four classes with one letter A: $M(1)=4$, $f(1; 5)=1/2$, and $u(1)=1/5$. 
In the second, there are four classes with two letters A: $M(2)=4$, $f(2; 5)=1/2$, 
and $u(2)=2/5$. By Eq.~(\ref{average}), $\langle u(5) \rangle =3/10$.

\begin{figure}[t!]
\includegraphics[width=8.0cm]{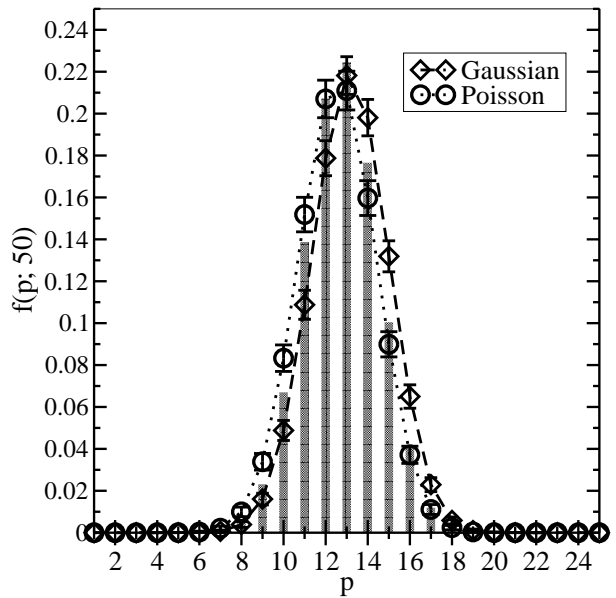}
\caption{\label{figure5} Probability distributions for $L=50$: theoretical 
(histogram), and from simulations with Poisson (circles) and 
Gaussian (diamonds) depositions. Error bars give one standard deviation from 
the mean time sequence at saturation. The frequencies were measured in  
$2048$ independent trials.}
\end{figure}

To find $f(p;L)$, one can exploit the recurrent structure of the binary tree for 
general $L$ in counting classes that contain  
the A-configuration at exactly $p$ sites \cite{KNR03}. Counting gives: $M=2^{L-2}$, 
$M(p)=(L-1)!/((2p-1)!(L-2p)!)$, and $p=1, 2, ..., [L/2]$ 
($[L/2]=L/2$ for even $L$; $[L/2]=(L-1)/2$ for odd $L$). Thus,
\begin{equation}
\label{distribution}
f(p;L) = \frac{1}{2^{L-2}} {L-1 \choose 2p-1} \, .
\end{equation}
In deriving $f(p;L)$ the underlying assumption is that at saturation any class of 
the entire surface configurations is generated with probability $1/M$. 
The theoretical $f(p;50)$ is compared with the simulated distributions in 
Fig.~\ref{figure5}. We considered three deposition types: the Poisson type, where  
$\eta_k$ was sampled from the Poisson distribution; 
Gaussian type, where $\eta_k$ was sampled from the Gaussian; and, the uniform deposition 
with $\eta_k$ being a uniform random deviate. In all cases, Eq.~(\ref{distribution}) 
is a very close approximation to time-averaged distributions measured in simulations.

\begin{figure}[t!]
\includegraphics[width=7.0cm]{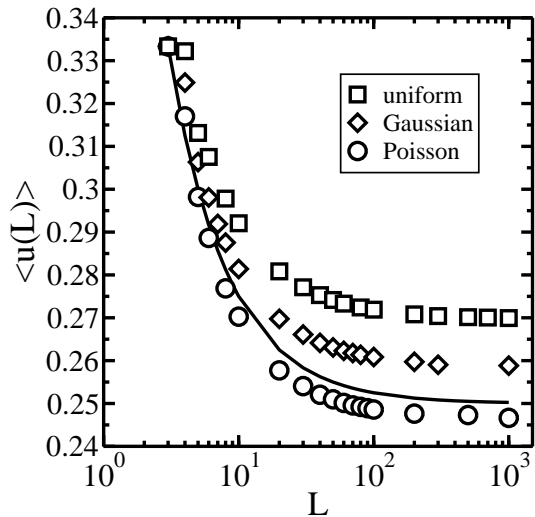}
\caption{\label{figure6} The steady-state mean density $\langle u(L) \rangle$ of local  
minima vs lattice size $L$. The curve is the analytical result 
$(L+1)/4L$. The symbols present  
simulations with the Poisson (circles), Gaussian (diamonds) and uniform (squares) 
depositions. The error bars are smaller than the symbol size.}
\end{figure}

Having Eq.~(\ref{distribution}), all moments of $f(p;L)$ can be computed 
exactly \cite{KNR03}, e.g., for $L \ge 4$ its variance is 
\begin{equation}
\label{variance}
\sigma ^2 (L) = \sum_{m} \left( m - \langle p (L) \rangle \right)^2 f(m;L) = 
\frac{L-1}{16} \, . 
\end{equation}
For $L \ge 3$ the mean number of local minima is
\begin{equation}
\label{mean_p}
\langle p (L) \rangle = \sum_{m} m f(m;L) =
\frac{L+1}{4} \, .
\end{equation}
The mean density of local minima, presented in Fig.~\ref{figure6}, is simply obtained 
from Eq.~(\ref{mean_p}) as $\langle u(L) \rangle = \langle p(L) \rangle /L$.

Suppose, the adopted growth rule is slightly modified so the deposition at a local 
minimum (site A) happens with probability $q_A$. Then, the new probability 
distribution $F(n; L, q_A)$ is the product of the probability $f(n_A; L)$ of drawing 
a surface that has exactly $n_A$ local minima, given by Eq.~(\ref{distribution}), 
and the probability $B(n; n_A, q_A)$ that $n$ depositions occur on this surface. 
The latter is given by binomial distribution
\begin{equation}
\label{binomial}
B(n; n_A, q_A) = {n_A \choose n} q_A^n \overline{q}_A ^{n_A - n} \, ,
\end{equation}
where first factor gives the number of ways in which $n$ depositions can be distributed 
among $n_A$ sites, and $\overline{q}_A = 1-q_A$ is the complementary probability of the 
no-deposition event at site A. In the old rule: 
$B(n; n_A, q_A=1) = \delta _{n,n_A}$ so, trivially, $F(n; L; q_A) = f(n_A; L)$.

Consider next a more complicated rule, where at the $t$-th deposition attempt, 
the deposition of $\eta _k(t)$ is not exclusively confined to local minimum 
but may also happen at other sites B, C or D with the corresponding 
probabilities $q_B$, $q_C$ and $q_D$, respectively. We assume 
now $q_A=1$. The probability 
$F(n;L) \equiv F(n; L; n_B, n_C, n_D; q_B, q_C, q_D)$ 
of having $n=n_A+n_B+n_C+n_D$ deposition sites 
in the interface is the product:
\begin{eqnarray}
\label{general}
F(n; L) & = &    
f(n_A; L) B(n_B; N_B, q_B) \nonumber \\
& \times &  B(n_C; N_C, q_C) B(n_D; N_D, q_D)\, . 
\end{eqnarray}
In Eq.~(\ref{general}), $f(n_A;L)$ is the probability of generating a surface with $n_A$ 
local minima, each of which is the deposition site; 
the binomial $B(n_B; N_B, q_B)$ is the probability of having $n_B$ depositions at $N_B$ 
sites in the local B-configuration on the generated surface; similarly, $B(n_C; N_C, q_C)$ 
and $B(n_D; N_D, q_D)$ are probabilities that $n_C$ and $n_D$ depositions occur at 
$N_C$ and $N_D$ sites in the local C- and D-configurations, respectively. For any surface, 
$N_A + N_B + N_C + N_D = L$. For surfaces growing on a closed ring of $L$ sites, 
the number $N_D$ of local maxima matches the number $N_A$ of local minima. This 
gives $N_B + N_C = L-2 n_A$. By Eq.~(\ref{average}), the mean 
number $\langle n(L) \rangle$ of deposition sites in the interface is
\begin{equation}
\label{sites}
\langle n(L) \rangle =  
\sum _{n_A=1} ^{ \left[ \frac{L}{2} \right] } 
\sum _{n_B=0} ^{N_B} \sum _{n_C=0} ^{N_C}\sum _{n_D=0} ^{n_A}  
 n F(n; L)\, . 
\end{equation}
The mean density of these sites is 
$\langle U(L) \rangle = \langle n(L) \rangle /L$.

In Eq.~(\ref{sites}), the summation over 
$n_D$ is easily performed: 
$\sum _{n_D=0}^{n_A}(n_A+n_B+n_C+n_D) B(n_D; N_D, q_D) = n_A (1+q_D) + n_B+n_C$. 
This gives
\begin{equation}
\label{partial}
\langle n(L) \rangle = \sum _{n_A=1} ^{ \left[ \frac{L}{2} \right] }
f(n_A; L) \left( n_A \left( 1 + q_D \right) + S_{B \vee C} \right) \, ,
\end{equation}
where $S_{B \vee C}$ 
is the mean number of deposition events in the set of sites other than A or D. 
The computation of $S_{B \vee C}$, though elementary, is a bit tricky since 
one must distinguish between depositions in the set of B-sites and depositions 
in the set of C-sites to avoid double counting:
\begin{eqnarray*}
S_{B \vee C} &=& 
\sum _{n_B=0} ^{N_B}B(n_B; N_B, q_B) \sum _{n_C=0} ^{N_C}B(n_C; N_C, q_C) \nonumber \\ 
& \times & (n_B+n_C) 
 = (L-2 n_A) q_B q_C \, . \nonumber
\end{eqnarray*}
Substitution to Eq.~(\ref{partial}) gives for $L \ge 3$:
\begin{equation}
\label{final}
\langle n(L) \rangle = \frac{1+q_D+ 2 \overline{Q}}{4} L + \frac{1+q_D - 2 \overline{Q}}{4} \, ,
\end{equation}
where $\overline{Q}=1-Q$, $Q=\overline{q}_B+\overline{q}_C- \overline{q}_B 
\overline{q}_C$.

\begin{figure}[t!]
\includegraphics[width=7.0cm]{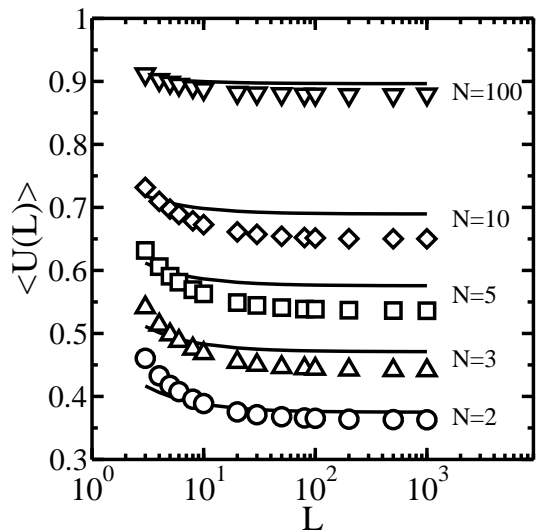}
\caption{\label{figure7} The steady-state mean density $\langle U(L) \rangle$ 
of deposition sites vs lattice size $L$ when $q$ depends on  
integer $N \ge 2$: analytical curves of Eq.~(\ref{density}) and  
the data measured in simulations with the Poisson deposition (symbols).}
\end{figure}

One application of Eq.~(\ref{final}) is the case when 
$\overline{q}_B= \overline{q}_C = q/2$ and $\overline{q}_D=q$, 
where the probability $q= \sqrt{2/N}$ 
is parametrized by an integer $N \ge 2$. Here, the mean density of lattice sites 
where the deposition takes place is 
\begin{equation}
\label{density}
\langle U(L) \rangle = \left( 1- \frac{q}{2} \right) 
\left( 1 - \frac{q}{4} \frac{L-1}{L} \right) \, .
\end{equation}
Predictions of Eq.~(\ref{density}), compared to simulations with the Poisson 
deposition in Fig.~\ref{figure7},    
are consistent with the fact that the approximation $f(n_A; L)$ gives 
slight overestimates of the density of local minima at the high-$L$ end 
(Fig.~\ref{figure6}). Nevertheless, the analytical $\langle U(L) \rangle$ 
reflects well the overall tendencies in the data. As $q \to 0$, $\langle U(L) \rangle$ 
approaches $1$. In this limit-case each attempted deposition event at each lattice site becomes 
certain. Such a case corresponds to the random deposition growth rule. 
The corresponding interface is entirely uncorrelated and never saturates. This is the 
asymptotic limit to which the curves in Fig.~\ref{figure7} converge when $N \to \infty$.

\section{OUTLOOK AND CONCLUSION \label{summary}}

In summary, we have introduced a new approach for the study of non-universal 
properties of surfaces. We showed the explicit 
connection between the interface morphology and the event statistics on the 
interface at saturation. The key concept here is the ability to build distinct 
equivalency classes of the entire surface configurations from its local-site 
configurations. In this construction, the growth rule defines the set of 
events that can be mapped in a unique way on the set of elementary local-site configurations.

Small differences between simulation results and the approximate analytical 
results come mainly from neglecting temporal correlations 
among $p$-groups of surface-configuration classes. 
These correlations are intrinsically present in the computation of 
averages over time series in simulations but they are absent in our derivation. They depend 
on the probability distribution from which the deposited 
random height increments are sampled. The proper handling of temporal correlations 
is essential in the microscopic description of the growth phase (before saturation), 
which is a challenging problem still open. 
Some insight to this issue 
may be found in recent works of Rikvold and Kolesik \cite{RK}, who used the concept 
of equivalency classes in connecting the interface  
microstructure with its mobility for Ising and solid-on-solid models with 
various stochastic dynamics.


Although the focus of the study presented here is on obtaining analytical 
results in the mean-field spirit, it is interesting to notice the 
sensitivity of simulation results to the probability distribution used 
to select random height increments. Negligible variations between the derived 
analytical probability distribution and simulated distributions with Gaussian, 
Poissonian and uniform increments demonstrate that the theoretical distribution 
presents well the statistics of events on the surface. 
Temporal correlations between interfaces generated via random 
deposition at local minima deserve further analytical studies.

In conclusion, we observe that our approach could be applied to 
a variety of other growth rules and models. The current applications to {\it 1D} 
problems are illustrative examples. The main advantage of the approach is that 
it enables one to compute analytically quantities that can be only estimated 
qualitatively by continuum methods.

\begin{acknowledgments}
This work is supported by NSF grants DMR-0113049 and DMR-01200310, and by 
the ERC Center for Computational Sciences at MSU. It used resources 
of the National Energy Research Scientific Computing Center, supported 
by the Office of Science of the US Department of Energy under contract No. DE-AC03-76SF00098.
\end{acknowledgments}

\end{document}